\documentclass[smallextended]{svjour3} 
\usepackage{bm}
\usepackage{graphicx} 
\usepackage{array} 
\usepackage{verbatim} 
\usepackage{subfig} 
\usepackage{tikz} 
\usetikzlibrary{shapes,snakes}
\usetikzlibrary{backgrounds,fit,decorations.pathreplacing} 
\newcommand{\ket}[1]{\ensuremath{\left|#1\right\rangle}} 
\usepackage{multirow}
\usepackage{marvosym}
\newcommand{\envelope}{(\raisebox{-.5pt}{\scalebox{1.45}{\Letter}}\kern-1.7pt)}
\usepackage{afterpage}
\usepackage{rotating}
\usepackage[nottoc,notlof,notlot]{tocbibind} 
\usepackage[titles,subfigure]{tocloft} 

\usepackage{fancyhdr} 
\usepackage{threeparttablex}
	\usepackage{amsmath,amssymb}
	\usepackage{afterpage}

	\usepackage{xcolor}
	
\newtheorem{thm}{Theorem}
		\newcommand{\braket}[2]{\ensuremath{\left\langle#1|#2\right\rangle}} 
		
\begin{document}
	\title{Multiple Network Alignment on Quantum Computers}
\author{Anmer~Daskin \and Ananth~Grama \and Sabre~Kais
}
\institute{A. Daskin -A. Grama \at
Department of Computer Science, Purdue University, West Lafayette, IN, 47907 USA
\and
S. Kais \at
Department of Chemistry, Department of Physics and Birck Nanotechnology Center,Purdue University,
West Lafayette, IN 47907 USA;
Qatar Environment and Energy Research Institute, Doha, Qatar
}
\date{Received: date / Accepted: date}
\maketitle
	\begin{abstract}

Comparative analyses of graph structured datasets underly diverse problems.
Examples of these problems include identification
of conserved functional components (biochemical interactions) across species,
structural similarity of large biomolecules, and recurring patterns of interactions in
social networks. 
A large class of such analyses methods quantify the topological similarity of nodes
across networks. The resulting correspondence of nodes across networks, also called
node alignment, can be used to identify invariant subgraphs across the input graphs.

Given $k$ graphs as input, alignment algorithms use topological information
to assign a similarity score to each $k$-tuple of nodes, with elements (nodes) drawn
from each of the input graphs. Nodes are considered similar if their neighbors are
also similar. An alternate, equivalent view of these network alignment algorithms is to
consider the Kronecker product of the input graphs, and to identify high-ranked nodes in the
Kronecker product graph. Conventional methods such as PageRank and HITS (Hypertext
Induced Topic Selection) can be used for this purpose. These methods typically
require computation of the principal eigenvector of a suitably modified Kronecker
product matrix of the input graphs. We adopt this alternate view of the problem to address
the problem of multiple network alignment. Using the phase estimation algorithm, we show that
the multiple network alignment problem can be efficiently solved on quantum computers.
We characterize the accuracy and performance of our method, and show that it can
deliver exponential speedups over conventional (non-quantum) methods.

	\end{abstract}
	\maketitle

\section{Introduction}

Recent developments have shown that quantum computers can efficiently solve diverse
important problems -- often delivering exponential seedups compared to their classical
counterparts \cite{Abrams,Papageorgiou}. Examples of such problems include finding
low energy states in lattice protein folding \cite{Perdomo2012}, simulation of chemical
dynamics \cite{Sanders2009algorithm,Sanders,Kassal2008polynomial}, calculation
of thermal rate constants \cite{New3}, Shor's factoring technique~\cite{Shor}, Grover's
search algorithm \cite{Grover} and
others \cite{Brown2010,New6,Kassal2011,Whaley2012,New1,New2,New4,Daskin2,Daskin2013,Childs2010}.  
The phase estimation algorithm \cite{Abrams,Kitaev}, used for finding eigenvalues
of a matrix, has been a key ingredient of many of these quantum algorithms
\cite{New1,New2,New4,Daskin2}. 

Graph structured datasets play an essential role in the representation of relationships
and interactions between entities. Comparative analyses of these datasets underly
diverse applications, including problems in chemoinformatics and bioinformatics.
A commonly used analysis technique aims to quantify the topological similarity of
nodes across a given set of graphs. Aligning nodes with high similarity, one may
identify approximate invariant subgraphs among the input graphs. The approximation,
in this case, is desirable, since it renders the underlying methods more robust to
noise in the input datasets. This paper focuses on the problem of multiple network
alignment. Specifically, it aims to develop quantum methods for computing the
topological similarity of nodes across a given set of $n$ graphs.

Techniques for quantifying topological similarity of nodes can be classified as
local or global. The former defines similarity on the basis of local neighborhoods
of nodes, while the latter uses the entire graph to compute similarity.
A commonly used global approach to computing node similarity uses the following
principle: two nodes are similar if their neighbors are similar. This principle
can be used to express the similarity matrix (a matrix whose $(i,j)$th element
corresponds to the similarity of node $i$ in the first graph with node $j$ in
the second) in an iterative form.

An alternate formulation of the same method
operates on the Kronecker product of input graphs. Given two graphs $G_1$ with
$r$ vertices and $G_2$ with $s$ vertices, with corresponding adjacency matrices
$A_1$ of dimension $r \times r$ and $A_2$ of dimension $s \times s$, the Kronecker
product of the graphs $G_1 \otimes G_2$ is computed through the adjacency
matrix $K_{12} = A_1 \otimes A_2$. One may view this Kronecker product matrix as the
adjacency matrix of the product graph. This graph has $r \times s$ vertices, labeled
$ij$. Vertex $ij$ has an edge to vertex $i'j'$ in the product graph iff there exists
an edge between vertices $i$ and $i'$ in the first graph {\em and} $j$ and $j'$ in
the second graph. In other words, the alignment of vertex $i$ in the first graph
to vertex $i'$ in the second graph and $j$ in the first graph to $j'$ in the
second graph is supported by the existence of an edge between the pair of aligned
vertices in {\em both} graphs. Now, consider powers of the matrix $K_{12}$. In
particular, the $ij$th element of the matrix $K_{12} \times K_{12}$ contains the
number of length two paths between vertices $i$ and $j$. This corresponds to the
number of {\em neighbor} alignments that support the $ij$th alignment. The argument can
be extended to higher powers of the matrix $K_{12}$. Stated alternately, the largest
entries in the higher powers of the matrix are the best aligned nodes. If the
matrix is suitably normalized (row-stochastic), then the principal eigenvector
reveals the strong alignments between the graphs $G_1$ and $G_2$. The argument
can be generalized beyond two input graphs to an arbitrary number of input
graphs, $G=\{G_1, \dots , G_n\}$. Note that the size of the Kronecker product graph grows
exponentially in the number of input graphs $n$. The principal eigenvector
of this graph reveals the strong complete vertex alignments by computing the alignment
score of each $n$-tuple \cite{Shahin2012,Kollias2011,Koyuturk2006,Singh2007}.

It is important to note that real-world graphs in bioinformatics and chemoinformatics
are typically large ($10^4$ vertices, $10^5$ edges and beyond, and 10 graphs and beyond).
The Kronecker product of these graphs can be viewed as combining local operators to
represent them as a global operator in quantum circuits. Furthermore, the sparsity of
these graphs renders then well-suited to efficient simulation on quantum
computers \cite{Aharonov2003}. Motivated by these considerations, we focus on
the problem of multiple graph alignment using the principal eigenvector formulation
of the implicit Kronecker product matrix. We show that for the types of the problems
where the dominant eigenvalue is known (1 in our case) and the corresponding eigenvector
is the solution to the problem, one can efficiently produce the solution as a
quantum state using the quantum phase estimation procedure. We show that
in the case of stochastic matrices, one can generate the solution as a quantum
state with the success probability 1. We present the quantum simulation algorithm,
quantify its cost, and show that in some cases, we can achieve exponential
improvement w.r.t. corresponding non-quantum methods.

In the following sections, we first briefly discuss the classical network alignment
method based on the PageRank and HITS ranking algorithms. We then describe the phase estimation
algorithm, and show how it can be adapted to the network alignment problem. Finally,
we discuss the representation of networks on quantum computers and the complexity
of our method to find the eigenvector (and consequently the node alignments).

\section{Principal Eigenvectors and Ranking Nodes in Networks}

Consider the problem of computing the principal eigenvector of a matrix using
a random walk on a suitably specified transition matrix. Assume, for generality that
the graph is directed; the case for undirected graphs is a special (symmetric) case
of the directed case.

Given a graph (in our case, the Kronecker product graph), the transition matrix $P$ is
constructed by setting value $p_{ij} = 1/deg(i)$, where $deg(i)$ is the out-degree of
node $i$, for each $j$ that node $i$ is connected to, and 0 otherwise. Note that this
matrix is not stochastic, since there may be nodes with no
out-edges; i.e., their transition probabilities sum to 0. A number of solutions have
been proposed to deal with this. Perhaps, the most commonly used is the PageRank 
formulation \cite{Page1999}, used in web ranking. Pagerank  deals with the problem by specifying a 
vector of a-priori probabilities to which a walker jumps to if there are no out edges
from a node. In this case, the new transition matrix is defined as $\hat{P}=P+\bold{dw}^T$,
where $\bold{w}$ is the a-priori probability vector, and elements of $\bold{d}$ are defined
to be 1 if $deg(i) = 0$, and 0 otherwise. Although this matrix is stochastic, it
is reducible; i.e., there may be multiple eigenvectors on the unit circle. To address
this, $\hat{P}$  is replaced by the matrix $\tilde{P}$ given by \cite{Brezinski2006}:
\begin{equation}
\tilde{P}=\alpha\hat{P}+(1-\alpha)E.
\end{equation}
Here, $\alpha \in [0,1]$ and $E = \bold{ev}^T$: $\bold{e}=[1,\dots,1]^T$,
and the vector $\bold{v}$ is called the personalization vector. It adds to all nodes a
new set of outgoing transitions with small probabilities.  The power iterations for
matrix $\tilde{P}$, $\bold{r} = \tilde{P}^T \bold{r}$, converge to a unique vector,
which is the eigenvector corresponding to the dominant eigenvalue of $\tilde{P}$,
which is 1.
 
\subsection{Example: PageRank inspired Protein-Protein Interaction (PPI) Network
Similarity}

\nocite{Waugh1954,Jones1996}  

Advances in high-throughput methodologies, supplemented with computational approaches
have resulted in large amounts of protein interaction data. This data is typically
represented as a protein-protein interaction (PPI) network -- an undirected graph,
$G(V, E)$, in which $V$ represents the set of proteins and edge $(v_i, v_j) \in E$
represents observed interaction between proteins $v_i$ and $v_j$ $\in V$. 
Comparative analyses of PPI networks of different species help in identifying conserved
functional components across species. The most common comparative analysis technique
is based on the alignment of PPI networks, where correspondences between nodes in
different networks are used to maximize an objective function. 
\cite{Shahin2012,Kollias2011,Koyuturk2006,Singh2007} 

Singh et al.\cite{Singh2007} proposed an iterative global algorithm called IsoRank,
in which the similarity measure between two nodes is determined by the similarity of
their neighbors. For two graphs, the  iterative relation for pairwise similarity
of nodes follows:
	  
	  \begin{equation}
	  \label{Eq:sing1}
	  R_{ij}= \displaystyle\sum_{u\in N(i)}\sum_{v\in N(j)}\frac{1}{|N(u)||N(v)|}R_{uv}.
	  \end{equation}
Here, $N(w)$ is the set of neighbors of node $w$; $|N(w)|$ is the size of this set;
$V_1$ and $V_2$ are the set of nodes for networks $G_1$ and $G_2$; and $i\in V_1$ and
$j\in V_2$. $R$ defines the functional similarity matrix whose stationary state is
used to find the solution for the alignment problem. Eq.(\ref{Eq:sing1}) can be
written in matrix form as:
	  \begin{equation}
	  \label{Eq:IsoRank1}
	  R=\tilde{A}R,
	  \end{equation}
where $\tilde{A}$ is a stochastic matrix constructed from the Kronecker product of the input
graphs with principal eigenvalue of one, and is defined as
$\tilde{A}=\tilde{A_1}\otimes \tilde{A_2}$. $\tilde{A_i}$ represents the modified adjacency
matrix for the graph $G_i$. The matrix $R$ is the stationary distribution of the
random walk on the Kronecker product graph.
Since $\tilde{A}$ has all positive entries, the infinite product of the matrix
will have a limit. The limiting matrix $R$ is the matrix with every row equaling
the left eigenvector associated with eigenvalue one. Therefore, it can be solved 
using the power method \cite{Golub}.  A-priori information regarding similarity of
nodes (for example, what proteins in one species are functionally related to proteins
in other species) can also be integrated into the iterative form. This a-priori
information may be derived from the Bit-Score of the BLAST (sequence)
alignments \cite{Tatusova1999}. The following equation integrates this a-priori
information in matrix $H$ as:
	\begin{equation}
	R=\alpha \tilde{A}R+(1-\alpha)H.
	\end{equation}
Here, entries of matrix $H$ define the Bit-Score between two nodes (proteins)
and $\alpha$ is a parameter that controls the weight of the network data relative
to the a-priori node-similarity data. This procedure can be written in an iterative
form as:
\begin{equation}
	R(k+1)=\alpha \tilde{A}R(k)+(1-\alpha)\bm{h}.
\end{equation}

This equation, in the limit, simplifies to the following \cite{Kollias2011}:
	\begin{equation}
	R(\infty )=
	(1-\alpha)\sum_{k=0}^{\infty}
	\alpha^k\tilde{A}^k\bm{h}.
	\end{equation}
The $R$ matrix can then be used in conjunction with a bipartite matching process to
identify a set of maximally aligned nodes across the networks. While our discussion has
been in the context of two networks, the procedure can be generalized to the
multiple network alignment problem \cite{Shahin2012,Singh2008,Liao2009}.
 
\subsection{Example: PageRank based Molecular Similarity}

Structurally similar chemical compounds generally exhibit similar properties.
Analyses of similarity of molecules plays an important role in infering properties
of compounds, and in designing new materials with desired characteristics~\cite{Bender2004}.
Graph kernels can be used to compute similarity of molecules. In this approach,
molecules are represented as undirected graphs, also called molecular graphs.
Vertices in these graphs correspond to atoms and edges correspond to covalent bonds.
In molecular graphs, vertices and edges are annotated with element and bond types.
These graphs can be large -- for instance, a muscle protein titin has $4.23\times10^5$ atoms.  
Different graph-based approaches have been proposed for molecular similarity.

Rupp et al.\cite{Rupp2007kernel} describe a technique based on iterative  graph
similarity.  Given molecular graphs $G_1$ and $G_2$ represented by their
adjacency matrices $A_1$ and $A_2$, the following update equation is
used for computing the pairwise similarity vector $x$:
\begin{equation}
\label{Eq:ms1}
\bm{x_{i+1}}=(A_1\otimes A_2)\bm{x_{i}}.
\end{equation}
In order to include more molecular graph properties (e.g. number of bonds), this formula is subsequently modified as:
\begin{equation}
\label{Eq:ms2}
\bm{x_{i+1}}=(1-\alpha)k_v+\alpha 
\times 
\text{max}_P P\bm{x_i},
\end{equation}
where $k_v$ is the vector of kernel functions between
vertices (similarity  functions measuring the similarity of pairs of atoms in graphs by including information on their certain substructures), and $P$ is a square matrix, compliant with the neighborhood structure, whose rows represent the possible neighbor assignments. 
For detailed explanation, refer to Rupp et al. \cite{Rupp2007kernel}.

\subsection{Ranking Nodes in a Graph: The HITS Algorithm}

Consider a directed graph $G(V,E)$ with $m$ vertices. Each node now has two
parameters -- an authority
weight and a hub weight. These weights are determined by the inherent quality
of the node and the number of edges to other authoritative nodes, respectively.
The HITS (Hypertext Induced Topic Selection) algorithm of Kleinberg \cite{Kleinberg1999}
associates a non-negative authority weight $a_i$ and a non-negative hub weight $h_i$ with
node $i$. Since the algorithm was originally proposed for ranking web pages, nodes
correspond to web pages and edges to hyperlinks. HITS computes numerical estimates
of hub and authority scores using an iterative procedure: if a node is pointed
to by many good hubs, its authority is increased in the next iteration. For
a node $i$, the value of $a_i$ is updated to be the sum of $h_i$ over all nodes that
point to $i$: 
\begin{equation}
 a_i =\sum_{j : j\rightarrow p}  h_j.
\end{equation}
The hub weight of the page is also increased in a similar way:
\begin{equation}
 h_i =\sum_{j : p\rightarrow j}  a_j.
\end{equation}
As a result, the update rules for the vectors of the authority weights and the
hub weights of the pages, respectively $\bm{a}$ and  $\bm{h}$, can be written as: 
\begin{equation}
\begin{split}
& \bm{a} = A^T \bm{h} =A^T A\bm{a} = (A^T A)\bm{a}\\
& \bm{h} = A\bm{a}= AA^T \bm{h} = (AA^T )\bm{h}.
\end{split}
\end{equation}
Vectors $\bm{a}$ and $\bm{h}$ converge to the principal eigenvectors of
$A^T A$ and $AA^T$, respectively. 

Lempel and Moran \cite{Lempel2000stochastic} modified the HITS algorithm to a
random walk on a graph: for the adjacency matrix $A$, a stochastic matrix $W_r$
is constructed by dividing each entry of $A$ by its row sum. Similarly, another
stochastic matrix $W_c$ is generated by dividing each entry of $A$ by its column
sum. Then, the iterations for the vectors of the authority and hub weights are
computed as defined \cite{Farahat2006authority}:
\begin{equation}
\begin{split}
& \bm{a} = W_c^T W_r\bm{a}\\
& \bm{h} = W_r^T W_c\bm{h}.
\end{split}
\end{equation}
Initially, $\bm{a}$ and $\bm{h}$ are unit vectors. Since $(W_c^T W_r)$
and $(W_cW_r^T)$ are stochastic matrices, $\bm{a}$ and $\bm{h}$ converge to
the principal eigenvectors of, respectively, $(W_c^T W_r)$ and $(W_cW_r^T)$ associated
with the eigenvalue 1.

\subsection{Network Similarity using the HITS Algorithm}

Blondel et al. \cite{Blondel2004measure} propose an iterative network similarity
algorithm based on the HITS method. For directed graphs $G_1$ and $G_2$ with
adjacency matrices $A_1$ and $A_2$, the iterative equation for the algorithm
is given by:
\begin{equation}
X_{k+1}=A_2X_kA_1^T+A_1X_kA_2^T,
\end{equation}
where $X_0$ is the matrix of all ones.
This equation is converted to the vector form:
\begin{equation}
	 \bm{x_{k+1}}=(A_1\otimes A_2+A_1^T\otimes A_2^T)\bm{x_k},
\end{equation}
where $\bm{x_k}$ is the vector form of $X_k$.

Having established the network similarity computation problem as one of computing
the principal eigenvector of a suitably defined matrix (with dominant eigenvalue 1),
we now focus on finding the steady state -- the principal eigenvector, of the
iterations defined in Sec.II and Sec.III on quantum computers.

\section{Phase Estimation Process and Eigenvector Generation}

The phase estimation algorithm \cite{Abrams,Kitaev} is a quantum algorithm for
estimating the eigenphase corresponding to a given approximate eigenvector of
a unitary matrix. For the eigenvalue equation:  $U\ket{\mu_j}=e^{i2\pi\phi_j}\ket{\mu_j}$,
it  finds the value of $\phi_j$ for a given approximate eigenvector $\ket{\mu_j}$.
The algorithm uses two quantum registers: $\ket{reg1}$ and $\ket{reg2}$.
While $\ket{reg1}$ is initially on zero state, $\ket{reg2}$ holds the eigenvector of
the unitary matrix. After putting  $\ket{reg1}$ into the superposition,
we apply  a sequence of operators, $U^{2^j}$, controlled by the $j$th qubit
of $\ket{reg1}$, to $\ket{reg2}$. This generates the Fourier transform of the
phase on $\ket{reg1}$. The application of the inverse quantum Fourier transform
makes  $\ket{reg1}$ hold the binary value of the phase. 
	
Assume that we have the operator $U=e^{i2\pi\tilde{A}}$,  where $\tilde{A}$
is the Kronecker product matrix for which we are trying to compute ranks.
The eigenvalues of the ranking matrix $\tilde{A}$ are known to be
$\lambda_1\leq \lambda_2 \leq \dots \leq \lambda_{N-1} < \lambda_N=1$,
associated with eigenvectors $\ket{\mu_1}, \dots ,\ket{\mu_N}$, where $N$ is
the size of $\tilde{A}$.  Consequently, in the ranking problem, we need
to find the eigenvector associated with the eigenvalue $\lambda_N=1$.

The above problem is, in some sense, the inverse of the standard phase estimation
algorithm. In the phase estimation process, instead of a particular eigenvector, if
the initial state of $\ket{reg2}$ is set to a superposition  of the eigenvectors
(not necessarily uniform), as  in Shor\rq{}s factoring algorithm\cite{Shor},
$\ket{reg1}$ holds the superposition of the eigenvalues of  $\tilde{A}$ in the final state.
Since the principal eigenvalue $\lambda_N$ is known, in the sate where we have
$\lambda_N$ on $\ket{reg1}$ (for the eigenvalue 1, the state where $\ket{reg1}=\ket0$),
we have the corresponding eigenvector on  $\ket{reg2}$, which is the solution
to the network alignment problem. 	 
\subsection{Steps of the Algorithm}
Here, we give the states in each step of the algorithm  applied to the ranking matrix $\tilde{A}$ with the eigenvalues
$\lambda_1\leq \lambda_2 \leq \dots \leq \lambda_{N-1} < \lambda_N=1$,
associated with eigenvectors $\ket{\mu_1}, \dots ,\ket{\mu_N}$:
\begin{enumerate}

\item Find the unitary operator $U=e^{i2\pi\tilde{A}}$, which  can be found easily if $\tilde{A}$ is sparse (see Section \ref{Section:Representation}).

\item Initialize the quantum registers \ket{reg1} and \ket{reg2} as
 $\ket{reg1}=\ket{0}$ and $\ket{reg2} = \ket{\mu}$, which is the superpostion of the eigenvectors (choosing $\ket{\mu}= H^\otimes\ket{0}$ makes the success probability of the algorithm equal to 1, see Section \ref{Section:SuccessProbability}.)

 \item Apply the quantum Fourier transform to \ket{reg1}, which produces the state:
 \begin{equation}
  \frac{1}{\sqrt{\kappa}}\sum_{j=0}^{\kappa-1}\ket{j}\ket{\mu}.
 \end{equation} 

 \item Apply $U^{2^j}$ controlled by the $j$th qubit of $\ket{reg1}$ to $\ket{reg2}$. If we only consider the $j$th qubit of \ket{reg1}, then the  following quantum state is obtained:
  \begin{equation} 
  \frac{1}{\sqrt{\kappa}}(\ket{0}\ket{\mu}+\ket{1}U^{2^j}\ket{\mu}).  
   \end{equation}
 $U^{2^{j}}\ket{\mu}$ generates a superposition of the eigenvectors with the coefficients determined by the eigenvalues.
  If we assume  $\ket{\mu}=\frac{1}{\sqrt{N}}\sum_i^N\ket{\mu_i}$, 
  then  for the eigenvector $\ket{\mu_i}$, 
  we have the coefficient $\lambda_i^{2^j}/\sqrt{N}$ in the state  $U^{2^{j}}\ket{\mu}$:
    \begin{equation} 
 \frac{1}{\sqrt{\kappa}}(\ket{0}\ket{\mu}+\ket{1}\frac{1}{\sqrt{N}}\sum_i^N U^{2^j}\ket{\mu_i})
  =\frac{1}{\sqrt{\kappa}}(\ket{0}\ket{\mu}+\ket{1}\frac{1}{\sqrt{N}}\sum_i^N \lambda_i^{2^j}\ket{\mu_i}).  
   \end{equation}
   Note that since the principal eigenvalue is 1, the largest coefficient is $\lambda_N^{2^j}/\sqrt{N}=1/\sqrt{N}$ and so  the dominant term in $U^{2^{j}}\ket{\mu}$ is the principal eigenvector $\ket{\mu_N}$.

  \item Apply the inverse Fourier transform to obtain the superposition of the binary form of the phases in $\ket{reg1}$.
  
  \item Finally, apply conditional measurement to \ket{reg1} to produce the eigenvector $\ket{\mu_N}$ on $\ket{reg2}$ corresponding the principal eigenvalue $\lambda_N=1$: i.e., if $\ket{reg1} = \ket{0}$, $\ket{reg2}=\ket{\mu_N}$. However, as discussed in Section \ref{Section:SuccessProbability} and Section \ref{Section:SuccessProbability2}, the success probability for the stochastic matrices is 1. Therefore, there is no need for conditional measurement. After Step 5 (the application of the inverse Fourier transform), \ket{reg1} = \ket{0}. Thus, \ket{reg2} holds the principal eigenvector.

\end{enumerate}

\subsection{Success Probability}
\label{Section:SuccessProbability}
\begin{figure}
\centering
	\includegraphics[width=3in]{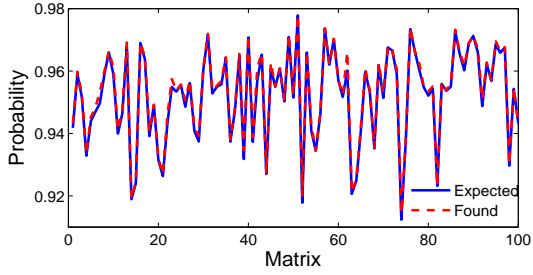}
	\caption{The success probabilities for 32x32 random symmetric positive matrices. }
		\label{fig:random100}
	\end{figure}

\begin{figure}
\centering
	\includegraphics[width=3in]{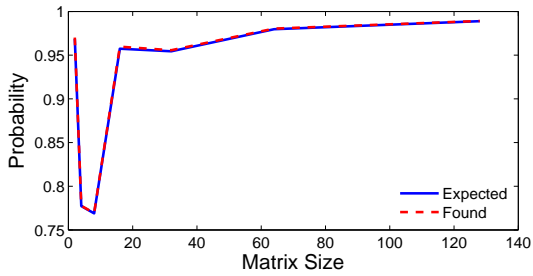}
	\caption{The success probabilities for random symmetric positive matrices of different sizes.}
		\label{fig:differentsizes}
	\end{figure}

The success probability of the algorithm is the  probability of observing
the principal eigenvalue on \ket{reg1}, which is related to the closeness of
the input to the principal eigenvector.	If we have  $\ket{\mu}=H^{\otimes n}\ket{0}$
as the initial input on \ket{reg2}, the amplitudes of the eigenvalues on
\ket{reg1} change depending on the closeness of the eigenvectors to this input.
We measure the closeness between an eigenvector and the input vector by using
the dot product of these vectors: i.e., the cosine of the angle between these vectors:
\begin{equation}
\left(\begin{matrix}
\braket{\mu_{1}}{ \mu}\\
\braket{\mu_{2}}{ \mu}\\
\vdots\\
\braket{\mu_{N}}{ \mu}\\
\end{matrix}\right)=
\label{eq:probabilites}
\left(\begin{matrix}
\frac{1}{\sqrt{N}}\sum_j\mu_{1j}\\
\frac{1}{\sqrt{N}}\sum_j\mu_{2j}\\
\vdots\\
\frac{1}{\sqrt{N}}\sum_j\mu_{Nj}\\
\end{matrix}\right)
=
\left(\begin{matrix}
\beta_1\\
\beta_2\\
\vdots\\
\beta_N
\end{matrix}\right)
\end{equation}
When the angle between two vectors is small, $\beta_i$s get larger.
The squares of the amplitudes in the above vector give us the success
probability for finding  an eigenvector on \ket{reg2} and the corresponding
eigenvalue on $\ket{reg1}$. For instance, the probability of observing the
eigenvector $\ket{\mu_N}$ on \ket{reg2} is $\beta_N^2$. Fig.\ref{fig:random100}
and Fig.\ref{fig:differentsizes} show the comparison of the expected
probabilities computed from Eq.(\ref{eq:probabilites}) with the probabilities
found in the phase estimation algorithm for random matrices with the dominant
eigenvalue 1. The following Perron-Frobenius theorem \cite{Meyer2000} 
provides the basis for comparing the  probability for the principal eigenvalue,
$\beta_N^2$, with the others:
\begin{thm}
For an irreducible non-negative  square matrix, the dominant eigenvalue is
positive and  has multiplicity one.  The eigenvector (unique up to scaling)
corresponding to this eigenvalue is also positive, and there are no other
non-negative eigenvectors for this matrix.
\end{thm}

Based on the above theorem, the vector $\ket{\mu_N}$ must be positive.
Therefore, the cosine of the angle between the input and the principal eigenvector,
$\beta_N=\braket{\mu_N}{\mu}= \frac{1}{\sqrt{N}}\sum_j\mu_{Nj}$, can be bounded by:
\begin{equation}
1\geq \beta_N>  \frac{1}{\sqrt{N}}.
\end{equation}
Here  $\beta_N$ is $\frac{1}{\sqrt{N}}$ only when  an element of the eigenvector
is one and the rest of the elements are zero. Since the principal eigenvector is
positive, and all the other eigenvectors include negative elements;
$\beta_N>|\beta_j|$, $1\leq j \leq N-1$. 
Fig.\ref{fig:random100} shows the success probabilities for a collection of 
$32\times32$ random matrices, while Fig.\ref{fig:differentsizes} shows success
probabilities for random matrices of various dimensions.
Random matrices used in these experiments are symmetric positive and generated
using the Wishard method \cite{Wishart1928,Mehta2004}. Here, one creates a
random matrix $X$ and uses the product $XX^T$ to generate a symmetric matrix.
We also scale the matrices so that the largest eigenvalue is one. 
We observe, in these experiments, that success probability is very high and grows
sharply with the size of the system. 

	
	 
\subsection{Success Probability in the Case of Stochastic Matrices}
\label{Section:SuccessProbability2}
For a row stochastic matrix $A$ with the eigenvalue $\lambda_k$, and the associated
eigenvector $\ket{\mu_k}$, the eigenvalue equation can be written as:
\begin{equation}
A\ket{\mu_k}-\lambda_k \ket{\mu_k}=0,
\end{equation}
or more explicitly,
\begin{equation}
\begin{split}
\left(\begin{matrix}
a_{11}& \dots &a_{1M}\\
\vdots& \dots &\vdots\\
a_{M1}&\dots& a_{MM}
\end{matrix}\right)\left(\begin{matrix}
\mu_{k1}\\
\vdots\\ \mu_{kM}
\end{matrix}\right)- \lambda_k \left(\begin{matrix}
\mu_{k1}\\
\vdots\\ \mu_{kM}
\end{matrix}\right)=\left(\begin{matrix}
0 \\
\vdots\\
0
\end{matrix} \right)\\
\left(\begin{matrix}
\mu_{k1}\bm{a_{1}}+& \dots &+\mu_{k1}\bm{a_{M}}
\end{matrix}\right)- \lambda_k \left(\begin{matrix}
\mu_{k1}\\
\vdots\\ \mu_{kM}
\end{matrix}\right)=\left(\begin{matrix}
0 \\
\vdots\\
0
\end{matrix} \right)
\end{split}
\end{equation}
Here, $\bm{a_j}$s are the column vectors. If we sum the rows on either side of the
equality, since $\sum_ja_{kj}=1$, we get the following:
\begin{equation}
\begin{split}
&\sum_j\mu_{kj}- \lambda_k \sum_j\mu_{kj}=0\\
&(1-\lambda_k)\sum_j\mu_{kj}=0
\end{split}
\end{equation}
Hence, for $k\neq N$; since $\lambda_k\neq 1$, $\beta_k=\sum_j\mu_{kj}=0$.
Since there is only one  nonzero value in the amplitude vector $\ket{\beta}$ defined in Eq.(\ref{eq:probabilites}),
$\beta_N=\sum_j\mu_{Nj}$ has to be 1. Thus, the success probability of the algorithm
is 1 for stochastic matrices, which is the case for the network alignment problem.

\subsection{Representation of Networks}
\label{Section:Representation}

A network of  $M_1$ nodes can be represented using an $M_1\times M_1$ adjacency
matrix $A_1$. In this case, $log(M_1)=m_1$ qubits are needed to represent this network
on a quantum computer. If there are $k$ such networks, the total number of
qubits for \ket{reg2} is $\sum_{j=1}^{k}m_j$. 

For each $U_j=e^{-iA_jt}$, one can find a quantum circuit design with
$O(M_j^2)$ number of operations. In network alignment, the Kronecker product of the
adjacency matrices is used: $A=A_1\otimes \dots \otimes A_k$. Hence,
$U=e^{-iAt}$ can be defined as $U=U_1 \otimes \dots \otimes U_k$. Therefore,
the total number of operations is the sum of operations needed for
each $U_j$, which is $\sum_{j=1}^{k}O(M_j^2)$. If the networks are of the same
size (in terms of number of vertices) $M$, then this sum becomes equal to $O(kM^2)$. 

However, these matrices are typically sparse and so can be simulated on quantum
computers  with an exponential speed-up \cite{Berry2007,Childs2011}.
The operator $U_j=e^{-iA_jt}$ is the same as the operator used in the continuous
time quantum  walk on a graph defined by the adjacency matrix $A_j$. It has been
shown that continuous time random walks can be performed on quantum computers
efficiently \cite{Aharonov2003}. The efficiency of the quantum walk has been studied
for different types of the graphs, and for a class of graphs it has been shown
that traversing is exponentially faster\cite{Kempe2003}. The exponential efficiency,
in general, can be observed when the circuit design for the adjacency matrix or
the Laplacian operator of a graph on $m_j$ qubits requires $O(poly(m_j))$ number
of one- and two-qubit operations. Lemma 1 in Ref.\cite{Aharonov2003}  states that
if $A_j$ is a row-sparse (the number of nonzero entries is bounded by $poly(m_j)$)
and $||A_j|| \leq poly(m_j)$, then $A_j$ is efficiently simulatable.
A Hamiltonian acting on  $m_j$ number of qubits is said to be efficiently
simulatable if there is a quantum circuit using $poly(m_j, t, 1/\epsilon )$
one- and two-qubit gates that approximates the evolution of $A_j$ for time $t$
with error at most $\epsilon$ \cite{Childs2011}.
Berry et al.\cite{Berry2007} present an algorithm that can simulate $A_j$ with
computational complexity bounded by $O((d^4\times  m_j^*||A_jt||)^{1+o(1)} )$,
where $d$ is the maximum degree of a vertex in the graph represented by $A_j$. 
This  complexity bound is further improved to $O(d^2 (d + m_j^* ) ||A_jt|| )^{1+o(1)}$
by Childs and Kothari \cite{Childs2011}.
Therefore, when $A_j$s are row-sparse, the implementation of the operator
$U_j=e^{-iA_jt}$ requires $O(poly(m_j))$ number of operations. Thus, the number
of operations for $U=e^{-iAt}$ is bounded by $O(k\times poly(m))$, where the
networks are assumed to have the same sizes. For dense  matrices, although
exponential efficiency has not been demonstrated, polynomial efficiency is achievable. 

Please note that ranking algorithms operate on a modified matrix $\tilde{A}$, instead
of $A$. These modifications can be mapped to the local and global rotation matrices,
which eases the difficulty of finding a circuit design. 


\subsection{Algorithmic Complexity}

The  complexity of the algorithm is dominated by  the complexity of the phase
estimation algorithm, which depends on the number of operations needed to implement
the adjacency matrices. Assuming all networks have the same size $M$, as shown
before, the total number of gates in the circuit implementing the evolution of
the product of the adjacency matrices $A$ is bounded by $O(kM^2)$. If there
are $\kappa$ qubits in the first register, then the phase estimation algorithm
requires $O(\kappa k M^2)$ operations excluding the quantum Fourier transform.
For a general case, this is more efficient  than the number of operations required
by the classical algorithms that are based on the power iterations \cite{Golub}. 
However, as shown in the previous section, when $A_j$s are row-sparse,
they can be efficiently simulatable. In this case, the computational complexity
is bounded by $O( poly(m)\kappa k)$, which gives us an exponential efficiency over
the classical case. Here, the eigenvector is produced as a quantum state.

\subsection{Precision and Eigenvalue Gap}

In our test cases, using six qubits in \ket{reg1} gave us enough precision to
get accurate results. However, when there are other eigenvalues close to one
or zero ($e^{i2\pi 0}=e^{i2\pi 1}$), then one must make the size of \ket{reg1}
sufficiently large to distinguish the principal eigenvalue from the rest. 
When the eigenvalue gap between the first and the second eigenvalues is small,
the algorithm may generate a vector combination of the eigenvectors corresponding
to the second and the first eigenvalues.

\section{Conclusion}

In this paper, we consider the problem of multiple network alignment. We formulate
the problem as one of ranking nodes of the Kronecker product graph of the input networks.
We use conventional PageRank \cite{Page1999} and HITS \cite{Kleinberg1999} algorithms for
computing the node rankings. We solve this problem on quantum computers by modifying the
well-known quantum phase estimation algorithm to generate the principal eigenvector
of a given operator. We discuss the computational complexity and show that our
algorithm has significantly lower computational complexity than classical algorithms. 
We also show that if the adjacency matrices for the networks are sparse,
exponential efficiency is possible. Our proposed framework provides a roadmap for solving
numerous other  problems that can be formulated as Markovian processes or ranking
problems, on quantum computers.


	\end{document}